\documentstyle[prd,aps,psfig]{revtex}
\draft
\newcommand{\be}{\begin{equation}}
\newcommand{\ee}{\end{equation}}
\newcommand{\bn}{\begin{eqnarray}}
\newcommand{\en}{\end{eqnarray}}
\newcommand{\bd}{\begin{displaymath}}
\newcommand{\ed}{\end{displaymath}}
\newcommand{\bnn}{\begin{eqnarray*}}
\newcommand{\enn}{\end{eqnarray*}}
\def\Ref#1{(\ref{#1})}
\def\Journal#1#2#3#4#5{#1,\ {\it #2} \ {\bf #3}, \ #4 \ (#5).}
\def\journal#1#2#3#4#5{#1,\ {\it #2} \ {\bf #3}, \ #4 \ (#5)}

\begin{document}
\title{ Possibility between earthquake and explosion seismogram 
differentiation by discrete stochastic non-Markov processes and local 
Hurst exponent analysis}
\author{Renat Yulmetyev$^{1}$ \thanks{e-mail:rmy@dtp.ksu.ras.ru},
Fail Gafarov$^{1}$\thanks{e-mail:gfm@dtp.ksu.ras.ru}, Peter H\" anggi$^{2}$, 
Raoul Nigmatullin$^{3}$ and Shamil Kayumov$^{3}$}  
\address{$^{1}$Department of Theoretical Physics, Kazan State
Pedagogical University, \\  Mezhlauk Street, 1 420021 Kazan, Russia
\\$^{2}$Department of Physics, University of Augsburg,
Universit\"atsstrasse 1,D-86135 
Augsburg, Germany\\
$^{3}$Department of Physics, Kazan State University,
Kremlevskaya Str. 18, 420018, 
Kazan, Russia}
%\date{}\
\maketitle 
\begin{abstract}
The basic purpose of the paper is to draw the attention of researchers 
to new 
possibilities  of differentiation of similar signals having different nature. 
One of examples of 
such kind of signals is presented by seismograms containing recordings of  
earthquakes 
(EQ's) and  technogenic explosions (TE's).
EQ's are among the most dramatic phenomena in nature. We propose 
here a discrete stochastic model for possible solution of a problem of strong 
EQ's 
forecasting and differentiation of TE's from the weak 
EQ's. Theoretical analysis is 
performed by two independent methods: with the use of
statistical theory of discrete 
non-Markov stochastic processes (Phys. Rev. E {\bf 62},6178 (2000)) and the 
local Hurst exponent. The following earth states have been considered among 
them: 
before (Ib) and during (I) strong EQ, during weak EQ 
(II) and during TE
 (III), and in a calm state of Earth core (IV). The
estimation 
of  states I, II and III has been  made on the particular examples of Turkey 
(1999) EQ 's, 
state IV  has been taken as an example of  underground TE.  Time recordings of 
seismic signals of the first four 
dynamic orthogonal collective variables, six various plane of phase
portrait of 
four dimensional phase space of orthogonal variables and the local Hurst
exponent 
have been calculated for the dynamic analysis of states of systems I-IV. The
analysis 
of statistical properties of seismic time series I -IV has been
realized with the help of a set of discrete  time-dependent functions
(time correlation function and first three memory functions),
their power spectra 
and first three points in statistical spectrum of non-Markovity. In all
systems studied we have found  out bizarre combination of the following 
spectral  characteristics: the fractal frequency spectra adjustable by 
phenomena of usual and restricted self-organized criticality,
spectra of a white and color noises and unusual alternation of Markov and 
non-Markov effects of long range memory, detected earlier 
(J.Phys.A {\bf 27}, 5363(1994)) only for hydrodynamic systems. Our research 
demonstrates that discrete non-Markov stochastic processes 
and long- range memory effects play a crucial role in the behavior of seismic
systems I -IV. The approaches, permitting to obtain
an algorithm of strong EQ's forecasting and to differentiate TE's from weak 
EQ's, 
have been developed.  
\end{abstract}
\section{Introduction}
The earthquakes are among the most mysterious and dramatic phenomena,
occurring in nature. As a result of sets and breakups of terrestrial
cortex or higher part of mantle over 
hundreds of thousands underground
pushes and fluctuations of the Earth  surface occur annually. They propagate 
over long
distances in the form of elastic seismic waves. Nearly thousands of
them are registered by  people. Annually nearly
hundred earthquakes cause 
catastrophic consequences: they affect big communities of people and
lead to great economical losses.

For the study of basic mechanisms underlying its nature modern numerical 
and statistical
methods are used now in modeling and understanding of the EQ
phenomenon.  
In papers \cite{Sor1},\cite{Sam} the modified renormalization
group theory with complex 
critical exponents has been studied for implications of EQ's
predictions. Long-periodic corrections found fit well the experimental data.
Then universal long-periodic corrections based on the modified
renormalization group theory have been used successfully
\cite{Anif} for possible 
predictions of the failure stress phenomenon foregoing  an EQ.
The failure stress data are in a good reliability with acoustic emission
measurements. In paper \cite{Sal} it has been shown that the log-periodic
corrections are of general nature, they are related to the discrete
scale invariance and complex fractal dimension. This idea has been checked
in \cite{Sor2}, \cite{Joh} for diffusion-limited-aggregate
clusters. The paradox of the expected time until the next EQ with an attempt of 
finding of 
acceptable distribution is discussed in \cite{Sor3}. New
explanation of Guttenberg-Richter power law related to the roughness of
the fractured solid surfaces has been outlined in \cite{Cha}.
Recent achievements 
and progress in understanding of the complex EQ phenomena from
different points of view are discussed in the recent review \cite{Sor4}. New
numerical methods like wavelets and multi-scale singular-spectrum analysis
in treatment of seismic data are considered in \cite{Yio}.

All these previous methods are developed for understanding the statistical and 
non-
stationary properties of EQ's and TE's. But in this paper we would like to 
demonstrate 
some possibilities related initially to differentiation of EQ's from TE's. This 
problem has not 
only scientific significance related with recognition of similar signals having 
physical origin. 
In recent time it was related also with some political problems associated with 
testing of 
nuclear explosions also.

Seismic data are an object of careful analysis and numerous
methods of their treatment are used especially for forecasting of EQ's with 
strong magnitudes. In spite of wide application of approaches
based on nonlinear dynamics methods, the Fourier and wavelet transformations 
etc., we have essential limitations, which narrow down
the range of applicability of the results obtained.  One of the main
difficulty is that the discrete character of the seismic signals
registration is not taken into account. Another factor, which should be
taken into account is related to the influence of local time effects. Alongside 
with of the
discreteness and the local behavior of the seismic signals considered here exist 
the third peculiarity, viz., the influence of
long-range memory effects.

In this paper, we present one of the possible solutions to forecasting strong 
EQ's 
and differentiating  TE's from weak EQ's. In this presentation we
consider of three important
factors for seismic signals registered in the form
of seismograms : discreteness, long- range memory  and local time behavior. Two 
new 
methods are used to  analyse these three
factors. The first one is based on seismograms
considered in the form of a discrete non-Markov statistical process along with
analysis of corresponding
phase portraits, memory functions and the non-Markovity parameters. The second 
method is based on the generalized conception of the
Hurst exponent. These methods have been used for careful analysis of seismic 
data and to differentiate
EQ's from  TE's . The results
obtained with the use of these methods are useful in recognition of
specific features of EQ's and TE's and can be
used for strong EQ's forecasting. 

The paper is organized as follows. In Section 2 we describe in 
brief the stochastic dynamic of time correlation in complex systems 
containing seismic signals by the discrete non-Markov kinetic
equations. The local fractal dimension and the corresponding Hurst
exponent are defined in the Section 3. The real data treatment with the
use of non-Markov conceptions has been realized in the Section 4. The
Section 5 contains some results obtained by the local Hurst exponent
method. The basic conclusions are discussed in the final Section 6. 
    
\section{The kinetic description of discrete non-Markov random processes}
In recent paper \cite{main} the statistical theory of discrete non-Markov random
processes has been developed. The basic elements, which are necessary for
understanding of other sections, are presented, in brief, here. In
accordance with the Refs. \cite{main}- \cite{Wehr} the  fluctuations 
of random
variable $\delta x_j =\delta x (T +j\tau), j=0,1, \cdots, N-1$ of a complex 
system 
can be represented as $k$-component state vector 
\be
{\bf A}_k^0(0)=(\delta x_0,\delta x_1,\delta x_2,\cdots,\delta
x_{k-1})=(\delta x(T),\delta x(T+\tau),\cdots,\delta
x(T+(k-1)\tau). \label{f1}
\ee 
Here  $\tau$ is a finite discretization time, $\delta x_j$ and  $<x>$
define fluctuations and mean
value correspondingly, $T$ is the beginning of the time series. 
They are defined by conventional relationships
\be
\delta x_j=x_j-<x>, \ <x>=\frac{1}{N} \sum_{j=0}^{N-1}x(T+j\tau). \nonumber
\ee
The set of state vectors forms a finite-dimensional Euclidean space, where the
scalar product of two vectors can be defined as
\be
  <{\bf A} \cdot {\bf B}>=\sum_{j=0}^{k-1} A_j B_j.
  \label{f1.1}
\ee  
The time dependence of the vector  ${\bf A}$ can be defined 
as result of discrete $m$-step shift
\bn
 {\bf A}_{m+k}^m(t)=\{\delta x_m,\delta x_{m+1},\delta x_{m+2},\cdots,\delta
  x_{m+k-1}\}  \nonumber \\
  =\{\delta x(T+m\tau),\delta x(T+(m+1)\tau), 
  \delta x(T+(m+2)\tau),\cdots,\delta x(T+(m+k-1)\tau\}, \label{f2}
\en
where  $t=m\tau$ and $\tau$ is a finite time step. 
Statistical parameters (absolute
and relative variances) can be expressed by means of the scalar product of two
vectors as following
\bn
\sigma^2&=&\frac{1}{N}<{\bf A}_N^0 \cdot {\bf
A}_N^0>=N^{-1}\{{\bf A}_N^0\}^2,\nonumber \\
\delta^2&=&\frac{<{\bf A}_N^0 \cdot {\bf A}_N^0>}{N <X>^2}.
\nonumber 
\en
We define the evolution operator for the description of evolution of the 
variables 
$\delta x_j$ as following
\be
\delta x_{j+1}(T+(j+1)\tau)=U(T+(j+1)\tau, T+j\tau)\delta x_j(T+j\tau)=U(\tau)
\delta x_j. 
\ee
One can write formally the discrete
equation of motion by the use of operator $U(\tau)$ in the form
\be
\frac{\Delta x(t)}{\Delta
t}=\frac{x(t+\tau)-x(t)}{\tau}=\frac{1}{\tau}\{U(t+\tau,t)-1\}x(t).
\ee
The normalized time correlation function (TCF) can be represented by 
Eqns. \Ref{f1} and \Ref {f2} (where $t = m\tau$ is 
current discrete time) as follows 
\be
a(t)=\frac{<{\bf A}_k^0 \cdot {\bf A}_{m+k}^m>}{<{\bf A}_k^0 \cdot
{\bf A}_k^0>}=\frac{<{\bf A}_k^0(0) \cdot {\bf A}_{m+k}^m(t)>}{<{\bf 
A}_k^0(0)^2>}.
\label{f3}
\ee
From the last equation \Ref {f3} one can see that  TCF  $a(t)$ is obtained by
projection of the final state vector ${\bf A}_{m+k}^m(t) $ \Ref
{f2}  on the initial state vector $ {\bf A}_k^0 (0)$.
Because of this property 
one can write the projection operator
in the linear space of state vectors 
\be
\Pi {\bf A}_{m+k}^m(t)={\bf A}_k^0(0)\frac{<{\bf A}_k^0(0) {\bf
A}_{m+k}^m(t)>}{<|{\bf A}_k^0(0)|^2>}={\bf A}_k^0(0) a(t). \label{f4}
\ee
The projection operator $\Pi$ has the following properties 
\be
\Pi =\frac {{| \bf A} _k^0 (0) > < {\bf A} _k^0 (0) |} {< | {\bf
A}_k^0 (0) | ^2 >}, ~~ \Pi^2 =\Pi, ~~ P=1-\Pi, ~~ P^2=P, ~~\Pi
P=0, P\Pi=0. \label {f5}
\ee 
The projection operators  $\Pi$
and $P$ are idempotent and mutually
complementary. Projector $\Pi$
 projects on the direction of initial state vector ${\bf A}_k^0 (0)$,
 while the projector
$P$ projects all vectors on the direction which is orthogonal to the
previous one. Let us apply the projection technique in the state vectors
space for deduction of the discrete finite-difference Liouville's equation
\bn
\frac{\Delta}{\Delta t}{\bf A}_{m+k}^m(t) =i \hat L (t, \tau) \label{f51}
{\bf A}_{m+k}^m (t), \\ \nonumber 
 \hat L(t,\tau)=(i \tau)^{-1}\{U(t+\tau,t)-1 \}. 
\en  
The first expression defines the Liouville's 
quasioperator $\hat L$ and the second
expression defines the evolution operator $U(t)$. Transferring from 
vectors  ${\bf A}_{m+k}^m$ to a scalar value of the 
TCF $a(t)$ by means of suitable projection
procedure one can obtain the closed 
finite-difference equation for the
initial TCF
\be
\frac {\Delta a(t)} {\Delta t} = \lambda_1  a(t) -\tau
\Lambda_1  \sum _ {j=0} ^ {m-1} M_1 (j\tau) a(t-j\tau) .\label {f6} 
\ee
 Here $\Lambda_{1}$ is the relaxation parameter while the frequency 
$\lambda_{1}$  defines the
eigen spectrum of the Liouville's quasioperator $\hat L$
in the following way
\be
\lambda_1=i \frac{<{\bf A}_k^0(0) \hat L
{\bf A}_k^0(0)>}{<|{\bf A}_k^0(0)|^2>},
~~ \Lambda_1=\frac{<{\bf
A}_k^0 \hat L_{12}\hat L_{21} {\bf A}_k^0(0)>}{<|{\bf
A}_k^0(0)|^2>}.\label{f7} 
\ee
The standard Liouville's equation is obtained easily
from Eqs. (6), (10) by means of the 
limit $\tau \to 0$. In this case the Liouville's quasioperator 
$\hat L$ is reduced to a
classical or quantum Liouvillian and is defined correspondingly by the
classical or quantum Hamiltonian of the system considered. The given approach
is true for non-Hamiltonian systems of arbitrary nature when the
Hamiltonian cannot be written together with conventional equations of
motion. The function $M_1 (j\tau)$ in the right hand side of Eqn. \Ref{f6}
is the first order memory function
\be
M_1(j\tau)=\frac{<{\bf A}_k^0(0) \hat L_{12}\{1+i\tau \hat
L_{22}\}^j \hat L_{21} {\bf A}_k^0(0)>}{<{\bf A}_k^0(0) \hat
L_{12} \hat L_{21} {\bf A}_k^0(0)>},~~M_1(0)=1.\label{f8}
\ee
Here we use the following notation for the matrix elements of the
splittable Liouvillian quasioperator
       $L_{i,j}=\Pi_i \hat L \Pi_j, \ i=1,2, \ \Pi_1=\Pi, \
       \Pi_2=P, \ \hat L_{11}=\Pi \hat L \Pi, \ \hat L_{12}=\Pi \hat
       L P, \ \hat L_{21}=P \hat L \Pi, \ \hat L_{22}=P \hat L
       P$.
Equation \Ref{f6} can be considered as the first equation
of the finite-difference kinetic equations chain with
memory for the discrete TCF $a(t)$. In paper \cite{main} it has been 
demonstrated that using  Gramm-Schmidt orthogonalization
 procedure one can define the
dynamic orthogonal variables ${\bf W}_n (t)$ by means of the 
following recurrence relationships
\be
{\bf W}_0={\bf A}_k^0(0),~~{\bf W}_1=\{i \hat
L-\lambda_1\}{\bf W}_0,
~~{\bf W}_n=\{i \hat
L-\lambda_{n-1}\}{\bf W}_{n-1}+\Lambda_{n-1} {\bf W}_{n-2},~~n>1.
\label{f9} 
\ee
Here we introduce the 
fundamental eigen values $\lambda_n$ and 
relaxation $\Lambda_n$ parameters as follows
\be
\lambda_n=i \frac{<{\bf W}_n \hat L {\bf W}_n>}{<|{\bf
W}_n|^2>},~~ 
\Lambda_n=-\frac {\langle  {\bf W}_{n-1}(i \hat L-\lambda_{n+1}){\bf W}_n
\rangle} {\langle | {\bf W}_{n-1} |^2 \rangle} \ . \label{f10}
\ee
Parameters $\lambda_n$ are very similar to the 
Lyapunov's exponents. Arbitrary orthogonal variables $ {\bf
W}_n$ can be expressed directly via the initial 
variable $ {\bf W}_0 ={\bf A}_k^0(0)$ by means of \Ref {f9}
in the generalized form 
\bn
{\bf W}_n=\left| \begin{array}{ccccc}
(i \hat L-\lambda_1) & \Lambda_1^{1/2} & 0 & \ldots & 0 \\ 
\Lambda_1^{1/2} & (i \hat L-\lambda_2) & \Lambda_2^{1/2} & \ldots & 0\\
0 & \Lambda_2^{1/2} &  (i \hat L-\lambda_3) & \ldots & 0 \\
0 & 0 & 0 & \ldots & (i \hat L-\lambda_{n-1})\\
\end{array} \right| {\bf W}_0. \label{f11}
\en
The physical sense of the new variables ${\bf W}_n$ can be
interpreted as follows. 
For example, the local density of fluctuations in the physics of
continuous media can be identified with the initial variable $ {\bf W}_0$.
In this case the fluctuations of the local current density, local energy 
density and
local energy current density can be associated with the dynamic variables
$ {\bf W}_n$ with numbers $n=1,2,3$, correspondingly.

One can relate to the set
of projection operators $\Pi_n$ 
to the set of orthogonal variables \Ref {f9}. The last ones project an arbitrary 
dynamic
variable (viz., a state vector) $Y$ on the corresponding initial state vector 
${\bf W}_n$
\bn
\Pi_n=\frac{| {\bf W}_n><{\bf W}_n^*|}{<|{\bf
W}_n|^2>},~~ \Pi_n^2=\Pi_n,~~ P_n=1-\Pi_n,~~ P_n^2=P_n,~~\Pi_n
P_n=0,\nonumber \\
\Pi_n\Pi_m=\delta_{n,m} \Pi_n,~~P_n P_m=\delta_{n,m}
P_n,~~ P_n\Pi_n=0.  \label{f12}
\en
Acting successively by projection operators $\Pi_n$ and $P_n$ on the
finite-difference equations \Ref{f51} for the normalized discrete memory
functions 
\be
 M_n(t)=\frac{<{\bf W}_n[1+i\tau \hat L_{22}^{(n)}]^m {\bf
 W}_n>}{<|{\bf W}_n(0)|^2>}  \label{f13} 
\ee           
one can obtain a chain of the coupled non-Markov
finite-difference kinetic equations of the following type
\be   
\frac{\Delta M_n(t)}{\Delta t}=\lambda_{n+1} M_n(t)-\tau
\Lambda_{n+1} \sum_{j=0}^{m-1} M_{n+1}(j\tau)M_n(t-j\tau) \ , \\
 \label{f14}
\ee 

Here $\lambda_n$ is the eigen value spectrum of Liouville's
operator $i \hat L$, while $\Lambda_n$ is 
the general relaxation parameters,                       
\bn
\lambda_n=i \frac{<{\bf W}_n^* L {\bf W}_n>}{<|{\bf W}_n|^2>}, \ \
\Lambda_n=-\frac {\langle  {\bf W}_{n-1}(i \hat L-\lambda_{n+1}){\bf W}_n
\rangle} {\langle | {\bf W}_{n-1} |^2 \rangle} \ \label{f10}, 
\nonumber
\en
which were defined before by relationships (15). 
One can consider the set of the functions $M_n (t)$
together with the initial TCF ($n=0$) 
\bn 
M_0(t)=a(t)=\frac{<{\bf A}_k^0(0) {\bf A}_{m+k}^m(t)>}{<|{\bf
A}_k^0(0)|^2>}, ~~t=m\tau,     \nonumber
\en
as  functions characterizing the
statistical memory of the complex system with discrete time. The initial TCF
$a(t)$ and the set $M_n(t)$ of discrete memory functions appearing from
Eqns.\Ref{f14} are playing an important role for the description of non-Markov 
and
long-range memory effects. Now it is convenient to rewrite the
set of Eqns. \Ref{f14} as the chain of the couplednon-Markov discrete 
equations for initial discrete TCF $a(t)$ ($t=m\tau$) and represent them
in the form  
\bn\frac{\Delta a(t)}{\Delta t}&=&\lambda_1 a(t)-\tau \Lambda_1 \sum_{j=0}^{m-1}
M_1(j\tau) a(t-j\tau), \nonumber \\
\frac{\Delta M_1(t)}{\Delta t}&=& \lambda_2 M_1(t)-\tau
\Lambda_2 \sum_{j=0}^{m-1} 
M_2(j\tau) M_1(t-j\tau), \nonumber \\
\frac{\Delta M_2(t)}{\Delta t}&=& \lambda_3 M_2(t)-\tau
\Lambda_3 \sum_{j=0}^{m-1} 
M_3(j\tau) M_2(t-j\tau).
\label{f15}
\en
The kinetic finite-difference Eqns. \Ref{f14} and \Ref{f15} are
analogous to the
well-known chain of kinetic equations of the Zwanzig'-Mori's type. These
equations are playing a fundamental role in the modern statistical
mechanics of nonequilibrium phenomena with continuous time. One can consider the 
kinetic
equations \Ref{f15} as a discrete-difference analogy of
hydrodynamic equations for physical phenomena with discrete
time. On the basis of the initial set of the experimental data one can find the 
set 
of orthogonal variables ${\bf W_n}$ in the following 
way 
\bn
\hat {\bf W}_0={\bf A}_k^0,~~\hat {\bf
W}_1=\left(\frac{\Delta}{\Delta t}- \lambda_1 \right){\bf A}_k^0 ,\nonumber \\
\hat {\bf W}_2=\left(\frac{\Delta}{\Delta t}-\lambda_2\right)
{\bf W}_1+\Lambda_1 {\bf 
A}_k^0=\{\left(\frac {\Delta}{\Delta t}\right)^2-
\frac{\Delta}{\Delta t} (\lambda_1+ \lambda_2) + \lambda_1
\lambda_2 + \Lambda_1 \}{\bf A}_k^0,\nonumber \\
\hat{\bf W}_3=
\left(\frac{\Delta}{\Delta t}- \lambda_3\right){\bf W_2} +
\Lambda_2 \left(\frac{\Delta}{\Delta t}-\lambda_1\right){\bf A}_k^0 .
\label{f16}
\en
It seems to us that one could suggest more physical 
interpretation of  different terms in the right side of the three Eqs.
(21). 
For example, term $\frac{\Delta A}{\Delta t}$ can be associated with 
dissipation,  term  
$\frac{\Delta^2 A}{\Delta t^2}$ is similar to inertia and term $\Lambda_1 
A(t)$ 
is related to restoring force. Then the third finite-difference derivative 
$\frac{\Delta^3 A}{\Delta t^3}$ is associated with the finite-difference form of 
the 
Abraham-Lorenz force corresponding to dissipation feedback due to radiative 
losses as seen from recent experimental evidence in frictional systems 
\cite{main}. 

In concrete applications it is necessary to take into account that the dimension 
of new state vectors
${\bf W}_n$  is gradually decreasing with the increase of the number $n$. 
If the initial vector ${\bf A}^0_k$ has dimension the  $k$ then the 
vectors ${\bf W}_1, \ {\bf W}_2$ and ${\bf W}_3$  will have the
dimensions $k-1$, $k-2$ and $k-3$, correspondingly.

Solving the chain of Eqns. \Ref {f14} under the
assumption that all $\lambda_s=0$ one can find recurrence
formulae for memory functions of arbitrary order in the following form
\bn
M_s(m\tau)=-\sum_{j=0}^{m-1}
M_s(j\tau)M_{s-1}((m+1-j)\tau)+\varepsilon_s^{-1}\{
M_{s-1}((m+1)\tau)-M_{s-1}((m+2) \tau)\}, \\ \nonumber
\varepsilon_s=\tau^2\Lambda_s,~s=1,2,3,\cdots   .  \label{f17}
\en
By analogy with \cite{main} it is convenient 
to define the
generalized non-Markov 
parameter for frequency-dependent case as follows
\bn
\epsilon_i(\omega)=\left \{ \frac{\mu_{i-1}(\omega)}
{\mu_i(\omega)} \right \} ^{\frac{1}{2}}, \label{f18}
\en
where $i=1,2,,,$ and $\mu_i(\omega)$ is the power spectrum of the i-th memory
function. It is convenient
to use this parameter for quantitative description of long-range memory
effects in the system considered together with memory functions defined above.

The set of new parameters describes the discrete structure of the system 
considered 
and allows to
extract additional information related to non-Markov properties
of the complex (non-Hamilton) systems .

\section{Local Hurst dimension analysis of seismic data}
{\bf The Hurst exponent.}
Typical seismic data are seismic waves registration
written in the form of vibrations of the Earth surface. Many observations
as seismograms lead to random series registrations: technogenic noises,
gravimetrical, economical, meteorological and other data. Some properties
of such random series can be characterized by the Hurst exponent H
\cite{Hur1}, \cite{Hur2}. Let
$\xi_i$ define the $i$-th value of the observable variable, $<\xi_\tau>$ 
define its mean
value on the segment containing $\tau$ registered points. For the cumulative
average value we have $X(t,\tau)=\sum_{i=1}^t(\xi_i-<\xi>_r)$. 
The range R for the given sampling of the random
series considered is defined as follows
\be
R(\tau)=max X(t, \tau)-min X(t, \tau), \label{s1}
\ee
at $1<t<\tau$, where $t$ is discrete time accepting integer values and
$\tau$ is a length of the time sampling considered.

Normalizing the range $R$ on the standard deviation $S$ for the chosen
sampling $\xi_i$
\be
S(\tau)=\left (\frac{1}{\tau} \sum_{i=1}^\tau \{ \xi(t)-\langle \xi 
\rangle_\tau\}^2
\right )^{1/2}
\ee
and analyzing the variations of the normalized range Hurst \cite{Hur1},
\cite{Hur2} 
obtained the following empirical relationship:
\be
\frac{R(\tau)}{S(\tau)}=\tau^H \label{s2},
\ee
where $R$ is the range, $S$ is the normalized variance and $H$
is the so-called 
Hurst exponent for the sampling of the given length $\xi$.
The value $H=0.5$ 
corresponds to the normal distribution sampling, other values correspond
to the various degrees of correlations, which can be interpreted in the
terms of persistent coefficient. One
can use the normalized range method for the definition of the Hurst exponent, 
but it 
works well for large samplings
containing 1000-10000 registered points. 

{\bf The calculation of the Hurst exponent for seismic data.}
 One can obtain easily the Hurst
exponent for long ($1000-10000$ registered points) samplings
\cite{Fed} by means of the method
of the normalized range ($R/S$ analysis). The Hurst 
exponent restoration accuracy calculated on the model data is located in
the interval (0.1-1\%). 
For example, if the model Hurst exponent was chosen
as $0.7$ then in the result of the $R/S$ analysis the restored value is
equaled to $0.69$ with the changeable third decimal point. The calculated
Hurst exponent for the initial seismic noise without an "event"
(earthquake/explosion) accepted the values $0.96-0.98$. The obtained values
show the high level of persistency and correlation. However, these values can be
referred to the whole series and cannot reflect the peculiarities of the event.
In other words, the values of the Hurst exponents calculated for the whole
series cannot provide information about possible EQ's/TE's, which can be 
characterized by other values of persistency. In this
situation it is necessary to generalize this parameter and define the
notion of local Hurst exponent.

{\bf The local Hurst exponent.} The generalized (local) Hurst exponent can be a
sensitive indicator, which gives  additional information about the 
regular component in the sampling considered. 
But the reason for changing of the Hurst exponent $H$ is not
only  the presence of the signal in the sampling considered but  slow
(for natural processes) variations of the correlated noise itself.
 
If one considers random series for a relatively long time it is logically 
appropriate 
to cut the series into short segments and calculate the Hurst exponent $H$ for
each of them. In such a manner, one can detect the variations of $H$ on time
or in some spatial coordinates. It is better to use the shortest
intervals possible for calculating  the local exponent $H(t)$.  
$Sufficient$ number of registered points can serve as a criterion for choosing 
the minimal
interval for such kind of statistical calculation of the local exponent
$H$. So by analogy with
conventional definition of the local temperature in statistical physics
one can generalize the conception of the Hurst exponent and use it for
short samplings.   
The reasons for changing of
the Hurst exponents can be the following: a) slow changing of  type of
correlations inside the noises; b) the presence of the regular signal inside the 
noises.  So, in concrete applications the local Hurst exponent can serve as a
$quantitative$ characteristic reflecting the fractal properties of the EQ or TE 
event. It is obvious that the usage of long intervals
($1000$ registered points and more) for the calculation of the local Hurst
exponents becomes useless and the important question is choosing the  acceptable 
interval for calculating this parameter with high accuracy.
The usual $R/S$ analysis does not give the acceptable accuracy for the local
Hurst exponent related to short samplings containing $100-120$ points. So it is
necessary to change the method of calculation of the local Hurst exponent for
short samplings.
The reliable calculation of the Hurst exponent averaged over short
samplings  turned out to be a nontrivial procedure and required 
elaboration of stable algorithms adjusted for averaging of short segments
of the given samplings.

We used another definition for the Hurst exponent \cite{Fed},
which turned out to be more effective for short samplings. 
The best results have been achieved in the usage of expression for the
normalized dispersion, which relates differences of a random function to 
retardation time $\tau$: 
\be
V(\tau)=\frac{\langle (B_H(t+\tau)-B_H(t))^2\rangle}{\langle
B_H^2(t)\rangle}=\tau^{2H},
\ee
here $B_H(t)=\sum_{i=1}^t \xi_i$, is an integral random function.
As a the result
of numerical experiments it has became possible to calculate the local Hurst
exponent with acceptable accuracy ($4-5$ decimal points) for samplings
containing about $80-120$ registered points. 
\section{NON-MARKOV DISCRETE ANALYSIS OF SEISMIC DATA}

Here we will apply the discrete non-Markov procedure developed in the previous 
Section
2 for the analysis of the real seismic data.  The basic problems, which 
we are trying to solve 
in  this analysis, are the following. The first problem 
relates to a possibility of seismic activity description by statistical
parameters and functions of  non-Markov nature. The second problem 
relates to distinctive parameters and functions for
differentiation of weak EQ's (with small magnitudes) from
TE's. The third problem is the most
important one and relates to strong EQ's forecasting. With this aim in mind we
analyzed three parts of the real seismogram: before the event
(EQ and TE), during the event and after the event. A typical
seismogram contains 4000 registered points. The complete analysis includes
the following information: phase portraits of junior dynamical
variables, power spectra of four junior memory functions and
three first points of statistical spectrum of non-Markovity parameter. We
took into account also the values of numerical parameters characterizing
the seismic activity. To analyze time functions we used also the
power spectra obtained  by the fast Fourier
transform. The complete analysis exhibits great variety of
data. 

We used 4 types of available experimental data courteously given by
Laboratory of Geophysics and Seismology (Amman, Jordan) for the following
seismic phenomena: strong EQ in Turkey (I) (summer 1999), a weak
local EQ in Jordan (II) (summer 1998). As a TE
we had the local underground explosion (III). The case (IV) corresponds to
the calm state of the Earth before the explosion. All data correspond to 
transverse seismic displacements. The real temporal step of
digitization $\tau$ 
between registered points of seismic activity has the following values, viz., 
$\tau=0,02 s$ for
the case I, and $\tau=0,01 s$  for the cases II-IV. The graphical
information is classified as follows: 

Figs.1-6 are referred to the case I;
Figs. 7,8 correspond to the cases II and III considered together; Figs.9,10 
illustrate the case IV. At first we consider the figures, which were obtained 
from the recordings
corresponding to the states defined as {\it before} and {\it during}
strong earthquake (EQ). Figs.1(a,b,c,d) 
(before EQ, state Ib) and Figs.1(e,f,g,h) (during EQ, state I) demonstrate the 
temporal
dynamics of four variables $W_0(t), W_1(t), W_2(t), W_3(t)$,
which were calculated in accordance with Eqns.
\Ref{f9}, \Ref{f11}. Let us note, that for convenience we use throughout initial 
variable 
$W_0(t)$ as a
dimensionless variable.
From these figures it follows that for variable $W_0$ the scale
difference achieves the value more than 2500 (compare Figs.1a
and 1e)! In comparison with the cases
(1b,c) the figures (1f,g) reveal the long-range and low frequency
oscillations for variables $W_1$ and $W_2$.  One can calculate phase portraits 
in four-dimensional space of the obtained 
four dynamical variables $W_0, W_1, W_2, W_3$  as well. Figs. 2 and 3 show six 
projections on various two-dimensional 
planes of states: before (Figs. 2a,b,c, Figs. 3a,b,c) and during (Figs. 2d,e,f, 
Figs.3d,e,f) EQ.

The phase portraits of the system analyzed demonstrate strong variations. The 
last arise owing to
the transformation of the
strained state of the Earth before the EQ to the state during the EQ. 
The most dramatic changes emerge in the phase plane ($W_1, W_0$) (see
Figs. 2a and 2d), plane $(W_2, W_0)$ (Figs. 2b and 2e), plane $(W_2,
W_1)$  (Figs. 3b and
3e) and$(W_3,
W_1)$ (Figs. 3a and
3d). One can notice strong qualitative variations in the structure of
phase portraits in the following planes: $(W_1, W_0), (W_2, W_0)$ and
$(W_2, W_1)$. Besides, we can see the quantitative change of space scales of 
dynamic 
orthogonal variables. 
The plane projection
$(W_0, W_1)$ remind a strange attractor. The changes of phase portraits in other 
planes are less noticeable (compare Figs. 2c and 2f,
Figs 3b, 3c,
3e and 3f). The weakest change is revealed in the phase portrait in the plane 
$(W_3, W_2)$. Probably, this phase portrait is less informative 
and encloses
quasi-invariant part of the total phase portrait.  Besides the spatial scales
change of the orthogonal variables $W_3$ and $W_2$ , other essential 
deformations of this phase portrait were not observed.
 
As it has been mentioned above it is
convenient to analyze the power spectra for comparison
of memory functions.  One can
divide these spectra into the following regions: ultra-low frequency range 
(ULFR), low
frequency range (LFR), middle frequency range (MFR) and high frequency
range (HFR). Figs. 4 and 5 demonstrate spectra of four memory functions 
$M_0, M_1, M_2, M_3$
before and during EQ. Before (Fig. 4a) and during EQ (Fig.4c) the power
spectrum of the initial TCF $M_0$  has a fractal form $1/\omega^\alpha$ 
in double-log scale. One can observe a peak in ULFR (Fig.4c) during EQ. The
power spectra of the first and second memory functions during
EQ (Figs. 4d and 5c) have also the fractal structure. The last one reflects the 
existence
of linear frequency 
dependence in double-log scale within the LFR, MFR and
HFR. The similar fractal-like 
behavior for the Turkish strong EQ is preserved for
the third memory function for the state during EQ (see Fig. 5d).

Figs. 6 demonstrates the power spectra of the first three points of the
statistical spectrum of non-Markovity parameter for the states
before (a, b, c) and during (e, f, g) strong the EQ. One can make
the following conclusions from Figs. 6 a-d. On the first
level of relaxation process (see, Fig. 6a) the strained state of
the Earth crust before EQ can be associated with Markov and
quasi-Markov behavior in ULFR and LFR, correspondingly. The
influence of non-Markov effects is reinforced in MFR with $5
\cdot 10^{-2} f.u.< \omega < 10^{-1} f.u.$,($1 f.u.=2\pi/\tau$). Strong non-
Markovity
of the processes considered for $\varepsilon_1(\omega)$ takes
place in HFR with $10^{-1} f.u. <\omega< 0.5 f.u.$. Simultaneously we have the 
numerical values
$\varepsilon_2(\omega), \varepsilon_3(\omega)\sim1$  in the whole frequency 
region(see, Figs.
6 b, c). But this behavior implies that strong
non-Markovity effects are observed in these cases.

The similar picture becomes unrecognizable for seismic state during the
strong EQ (see, Figs. 6 d-f). Firstly, it is immediately
obvious that $\varepsilon_1(\omega)\sim1$ on first relaxation level.

Secondly, the second and third relaxation levels are non-Markovian
(see, Figs. 6 e, f). Thus, the behavior of seismic signals
during the strong EQ is characterized by strong non-Markovity on the
whole frequency region. 

Figs. 7 depicts power spectra of MF $M_0$ and
$M_1$ for seismic states II and III. Figs. 8 shows power spectra of the first
three points of non-Markovity parameter $\varepsilon_i(\omega),
i=1, 2, 3.$ The preliminary results suggest that there is
remarkable difference between weak EQ's and TE's especially in the area of low 
frequencies.

The analysis of the phase portrait for weak EQ's and underground TE's lead to 
the 
following conclusions. Firstly, these portraits cannot be 
differentiated. It can be seen from the range of spatial scales of the dynamical 
variables $W_i$ and
$W_j$ and
from  the analysis of the phase points distribution forms. Secondly, it is
necessary to remark some peculiarities in power spectra of 
$\mu_i(\omega)$, $i=0,1,2,3$ (see, Figs.8 ) for the 
cases II and III. All these spectra have distinctive
similarities for the memory functions  $M_i(t)$  with numbers $i=0,1,2$ 
and $3$. The
character and the form of the spectra considered for the cases II and III
are very similar to each other. 
The same similarity is observed for the three non-Markovity parameters 
$\epsilon_i(\omega)$, $i=1,2,3$ (see, Figs. 8).

Nevertheless the analysis of the power frequency spectra
allows to extract distinctive specific features between the weak EQ's
and the TE's. Such quantitative criteria can be associated with frequency 
spectra of 
memory functions
$\mu_i(\omega)$ characterizing the long-range memory effects in seismic 
activity.
This new criterion allows to tell definitely a weak EQ from a
TE, viz., to differentiate case II from case III. 

A close examination of Figs. 8a and 8d shows that this
distinction appear in frequency behavior of the first point of
non-Markovity parameter $\varepsilon_1(\omega)$ close to the zero
point $\omega=0$. Specifically, the ratio of values $\varepsilon_1(0)$ for
weak EQ and TE equal
$\varepsilon_1^{II}(0)/\varepsilon_1^{III}(0)=0.92/0.57=1.61$.
 
Let us to analyze the results of seismic activity characterizing the
calm Earth state. Figs. 9,10 present the results of this analysis.
They will be useful for the comparison with the results obtained for EQ's and
TE's. 
The projections
of the phase cloud on all six planes ($W_i, W_j$), $i \not=j$ exhibit 
approximately the similar distribution of phase points. 
The power spectra for the memory functions with the same parity (see,
Figs.9) have a similar form. For example, for even order functions
$\mu_0(\omega)$ and $\mu_2(\omega)$ one
can notice  sharp peaks near the frequency $0.2 \ f.u.$ (see, Figs. 9a,c). 
In the spectrum of the senior function $\mu_2(\omega)$ (see Fig. 9c) 
additional peaks
in HFR appears. One can notice two groups of characteristic peaks near $0.2 
f.u.$ and
$0.4 f.u$ in odd memory functions 
 $\mu_1(\omega)$ and $\mu_3(\omega)$ (Figs.
9b, d). With the increase of order of the memory function the pumping over
effect of peak intensities from the MFR to the HFR takes place. The
frequency behavior of the three points of non-Markov parameters 
$\epsilon_1(\omega),\ \epsilon_2(\omega)$ and  $\epsilon_3(\omega)$  appeared to 
be practically the same. The behavior of the functions 
$\epsilon_i(\omega)$
exhibits the typical non-Markov character with small oscillations of random 
nature at LFR. The spectral characteristics of the system IV are very useful in 
comparison to the results obtained for the system I
(before strong EQ). 

Our observation shows that the zero point values of
non-Markovity parameters for calm Earth state are equal
$\varepsilon_1^{IV}(0):\varepsilon_2^{IV}(0):\varepsilon_3^{IV}(0)
\approx 4.99:0.947:0.861$. These values are convenient for the
comparison with similar values for the Earth seismic state
before  the strong EQ:
$\varepsilon_1^{I}(0):\varepsilon_2^{I}(0):\varepsilon_3^{I}(0) 
\approx 214.3:0.624:0.727$. The change
of ratio of the two first non-Markovity parameters
$\varepsilon_1(0)/\varepsilon_2(0)$ is particularly striking . This ratio is 
equal to
5.27 for the
calm Earth state, then it comes into particular prominence for the 
state before strong EQ:
$\varepsilon_1^{I}(0)/\varepsilon_2^{I}(0)\approx 343.4$. Thus,
this ratio changes approximately in 60 times! Hence, the behavior of
this numerical parameter is operable as a reliable diagnostic tool for  the 
strong EQ  
prediction.  The foregoing proves that the indicated value drastically increases 
in 
process of nearing to strong EQ.

Finishing this Section, we give some preliminary suggestions
relating to the strong EQs forecasting. They are related in comparison of
frequency spectra obtained for the calm Earth (Figs. 11,12) and seismic
activity data registered {\it before} a strong EQ (see, Figs. 2a,b,c;
3a,b,c; 4a,b; 5a,b and 6a,b). The comparison of the phase portraits demonstrates
the following peculiarities. In the phase portraits calculated for the senior
dynamical variables $(W_2,W_1)$, $(W_3,W_1)$ and $ (W_3,W_2)$ 
obtained for cases I and IV the
distinctions are not noticeable (see Figs. 3a,b). These distinctions become 
noticeable in the phase portraits of junior
variables $(W_1,W_0)$, $(W_2, W_0)$ and $(W_3,W_0)$ 
(see, Figs. 2a,b,c). One can observe a gradual stratification of the phase 
clouds 
with the 
growth of elastic deformations before the
strong EQ. It is necessary to recall the double
frequency difference for systems I and IV when comparing the frequency plots.
The dependence $\mu_0(\omega)$, $\mu_1(\omega)$, $\mu_2(\omega)$
and $\mu_3(\omega)$ for systems I and IV (see Figs. 4a, b; 5a,b; and 14 a-d)
is approximately similar, and qualitative difference is not
noticeable. One can notice some visual difference only for two spectra:
for the third memory function spectrum  $\mu_3(\omega)$  
and for the ULFR of the memory
function $\mu_0(\omega)$. So the power spectra 
of memory functions can be
used for the strong EQ forecasting. One can notice the similar changes in
the behavior of the functions $\epsilon_1(\omega)$, $\epsilon_2(\omega)$
and  $\epsilon_3(\omega)$ (see, Figs.6 a,b,c and 10 a,b,c). So one
can conclude that 
careful investigations of frequency behavior of memory functions $\mu_i(\omega)$   
and
functions $\epsilon_i(\omega)$ describing the statistical non-Markovity 
parameters  provide an accurate quantitative method of the strong EQ's 
forecasting. It is necessary to investigate
carefully the power spectra with the accurate localization of an
object and source, generating seismic signals, for
further elaboration of this method.

For more complete understanding of non-Markov properties of seismic signals we 
give  some kinetic parameters of our theory in Tables I-III. In Table 
I the full sets of kinetic parameters describing
non-Markov stochastic processes in five various seismic states have been 
presented:
before strong EQ(Ib), during strong EQ (I), during weak EQ (II), during TE (III) 
and for the  calm Earth 
state (IV). The data cited
in this Table are indicative of non-equilibrium properties
(parameters $\lambda_1, \lambda_2$ and $\lambda_3$), long-range
memory effects (parameters $\Lambda_1$ and $\Lambda_2$) and
non-Markov peculiarities (parameters $\varepsilon_1(0),
\varepsilon_2(0)$ and $\varepsilon_3(0)$). The differences under
observation for various seismic states are sufficient to
allow definite conclusions. 

For purposes of clarity, Table II illustrates the of comparison of
specific kinetic non-Markov parameters for two seismic state:
before strong EQ(Ib) and calm Earth states (IV). As will be seen from
Table II differences of parameters for these two states vary
within a broad range: from 2.8 (parameter $\Lambda_2$) to 44.0
(for parameter $\varepsilon_1(0)$). Similarly, Table III
contains comparison data for the other two seismic states: during
weak EQ(II) and during underground TE(IV).
Differences of parameters in this case are established within
more narrow limits: from 2.486 (for parameter $\Lambda_1$) to
1.614 (for parameter $\varepsilon_1(0)$). 

Thus, the existence of discreteness 
and long-range memory in behavior of seismic signals opens up new fields of use 
in the analysis of the Earth seismic activity. We can state with
assurance that the differences under observation favor the view
that the non-Markov parameters of our theory will be available for
strong EQ's forecasting and  differentiation of TE's  from weak EQ's.

\section{LOCAL HURST EXPONENT CALCULATIONS FOR AVAILABLE 
EARTHQUAKES AND TECHNOGENIC EXPLOSIONS DATA}
Available data for the calculation of the local Hurst exponents contains
3000-5000 registered points describing the visible part of a wavelet. This
number of the recorded points allows to use the procedure of the local
Hurst exponent $H(t)$ calculation. For the realization of the procedure
described in the previous Section 3 it is necessary to divide the whole
sampling containing 25000 points into small intervals of 100-200 points,
where the local Hurst exponent is supposed to be constant. On Fig. 11 we
show a typical plot of the function $H(t)$ calculated for a typical EQ. 
The same features of $H(t)$ behavior are conserved for wide class of
available weak EQ's.
Then we obtained the calculated values of the local Hurst exponents $H(t)$ for
available EQ's and explosions. Figs. 12 exhibits the typical behavior of
these functions. The sharp decreasing $(0.1)$ of the local Hurst function during
"an event" is typical for explosions. Then the values of the function 
$H(t)$
are relaxing slowly to their initial values. For EQ's one can notice more
gradual change of $H(t)$ before the event. The relaxation of $H(t)$ starts
from  higher  $(0.85)$ values and it comes back faster to its initial values 
in
comparison with explosions. Such behavior is preserved for weak
signals, when the ratio $S/N$  decreases. For these cases the criterion of EQ
 or TE distinction is related to the amplitude of the
Hurst exponent change during the analyzed event. It is necessary to
increase the number of registered points per unit of real time in order to 
obtain  a more distinctive picture, which can be more useful 
in differentiation of these events. It is
related to the fact that the sensitivity of correlations of a random
fractal value changing is associated with the lower temporal limit of 
the corresponding measurements. The smoothed change of $H(t)$ obtained for EQ's
opens a possibility of more accurate registration of $H(t)$ before the visual
wavelet of EQ's.
\section{Conclusion}      
We want to stress here again that these presented method have been applied 
successfully  for differentiation of EQ's from TE's. We hope that the results of 
this analysis 
can be applied to a set of phenomena related with differentiation of similar 
signals of 
different nature. In the result of this analysis we received a new possibility 
to forecast of 
strong EQ's approaching,analyzing only seismograms recorded for transverse 
seismic 
waves. Secondly, we received sufficient amount of information for the definite 
differentiation of weak EQ's from TE's.

In this paper we have presented the results of application of two new methods 
for
the study of dynamic, kinetic and spectral properties of seismic signals 
depicting EQ's and TE's  modulation. 
By the discrete non-Markov stochastic processes 
and the local  Hurst exponent analysis we have found explicitly some features of 
several different states of the Earth crust: states  of the
the Earth {\it before} and 
{\it during} strong and weak EQ's, during TE's. 
The used methods allow to present the seismogram analyzed in the form 
of set non-Markov variables and parameters. They contain great amount of the 
qualitative
and quantitative information about seismic activity. 
 
The dynamic information is contained in time  
recordings of  new  orthogonal dynamic variables, different  plane  
projections of the multidimensional phase portrait and the time dependence 
of local Hurst exponent. The information on the kinetic, spectral and
statistical 
properties of the system is expressed through time  dependence of the initial 
TCF, memory functions of junior orders, their power
and frequency spectra of the first three points of statistical spectrum 
of non-Markovity parameter.

The main advantage of our two new methods is a great amount of supplementary 
information 
about the properties of seismic signals 
makes.
 The problem is its correct application. What kind 
of possibilities can one expect? It is possible to answer as follows. 
Firstly, our preliminary study, convincingly demonstrates
that   the relevant and valuable information on non-Markov and discrete 
properties 
of the system considered is contained in seismic signals. In
all the studied systems (I- IV) 
we have found out unique manifestations of Markov, quasi-Markov  
and non-Markov processes on the particular behavior of the signals in a broad 
range of frequencies.

The similar results cannot be obtained, on principle, by other methods used in 
the
analysis
of seismic activity.
 
Secondly, in the non-linear non-Markov 
characteristics  some of well-known spectral effects are evident. Among them the 
following effects are exhibited noticeably: fractal 
spectra 
with an exponential function $ \omega ^ {-\alpha} $, which are 
connected to the phenomenon of usual (SOC) and restricted (RSOC) self-organized
criticality [15,16], behavior of some frequency spectra in the form of
white 
and color noises. Thirdly, the frequency spectra introduced above are
characterized 
by the particular alternation of Markov (fractal) and non-Markov spectra (such 
as
color or white noises). The similar alternation resembles
in particular the peculiar 
alternation of effects of a non-Markov and non-Markov behavior for
hydrodynamic systems in the statistical physics of condensed matter 
detected 
in papers [17], [18] for the 
first time. The fine specification of such alternation appears
essentially 
different for studied states I-IV. These features allow to view optimistically   
the solution of the problem of forecasting strong EQ's and
differentiation 
 TE's from  weak EQ's.

\section{ACKNOWLEDGEMENTS}
We would like to thank our Referee for careful reading of the initial version of 
our manuscript presented, valuable remarks and well-wishing comments, which 
undoubtedly improved the quality of the paper. One of the authors (R.R.N) 
gratefully acknowledges Dr. Tawfiq Ch.
Al-Yazjeen (Laboratory of Geophysics and Seismology, Amman, Jordan) for
making possible access to some seismology data for their subsequent
analysis. Authors thank Dr. L.O.Svirina for technical assistance.

This work partially supported by 
Russian Humanitarian Science Fund (Grant N 00-06-00005a) and 
NIOKR RT Foundation (Grant N 14-78/2000).

\newpage
\section{Figure captions}

Figure 1. The temporal dynamics of the first four dynamic variables
$W_0(t)$, $W_1(t)$, $W_2(t)$, $W_3(t)$:
a,b,c,d - before strong EQ, e,f,g,h - during of strong EQ. During strong EQ 
fluctuation scale increase drastically. It makes up: $2.5*10^3$ for initial 
variable $W_0(t)$,$10^2 $ for the first orthogonal variable $W_1(t)$,10 for 
$W_2(t)$ 
and 2 for  $W_3(t)$.  The existent trend vanishes gradually at transition from 
the
initial variable $W_0(t)$ to the third orthogonal variable $W_3(t)$. The 
fluctuation 
scale decreases sharply during the strong EQ.

Figure 2. The phase portrait projections on the planes of orthogonal variables
$W_0, W_1$ (a), $W_0,W_2$ (b), $W_0,W_3$ (c)-
before the strong EQ(Ib) and $W_0,W_1$ (d),$W_0,W_2$ (e), $W_0,W_3$ (f)
during  the strong EQ (I). The sharp difference is distinct
for seismic states Ib and I. The randomization of the phase portrait for state I 
begins from plane $W_0,W_2$. Together with the difference of the scale of 
fluctuation, 
one can observe the asymmetric distribution of phase clouds everywhere.

Figure 3. The phase portrait projections on the planes of orthogonal variables
$W_1, W_2$ (a), $W_1,W_3$ (b), $W_2,W_3$ (c)-
before the strong EQ (Ib) and on planes $W_1,W_2$ (d),$W_1,W_3$ (e), $W_2,W_3$ 
(f)
during of strong EQ (I). All phase clouds for seismic 
state 
Ib are symmetrical as opposed to Figs.2. Sharply marked asymmetry and 
stratification of phase clouds, what resembles known situation for myocardial 
infarction in cardiology, are observed for state I (d,e, and f).

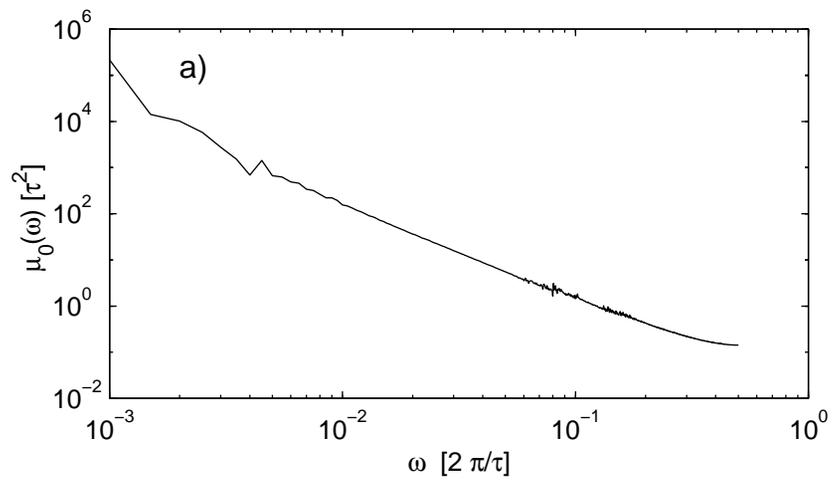
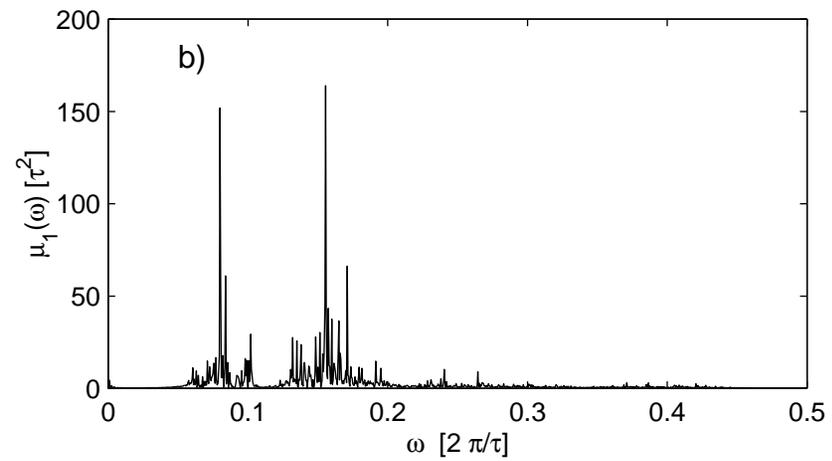
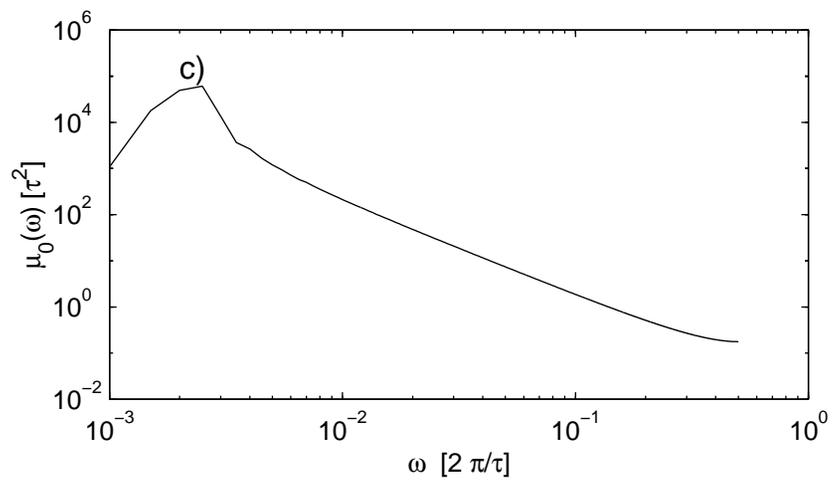
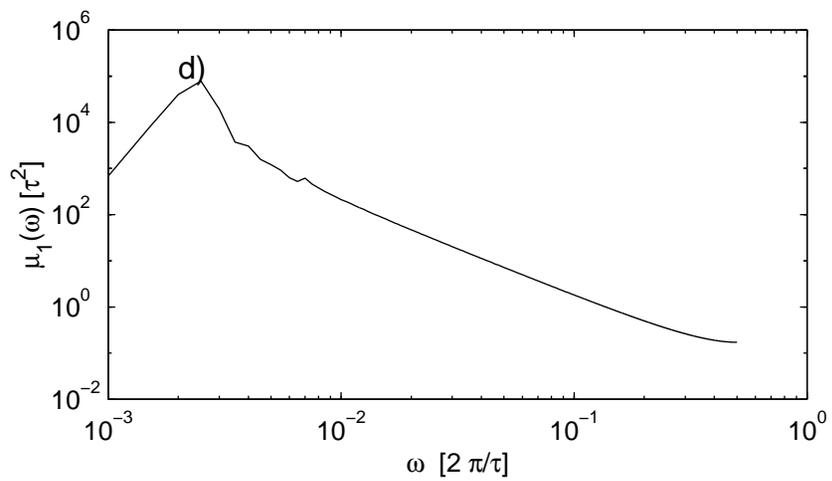
Figure 4. The power spectra of the two first memory functions $\mu_0$ and 
$\mu_1$
: (a,b) -before the strong EQ(Ib),  (c,d)- during of the strong EQ (I).For the 
cases a,c and d we observe fractality and self-organized criticality (SOC). SOC 
exists for the whole frequency range for state Ib. However, we observe 
restricted SOC at c and d cases only in frequency range down to $ 2.5*10^{-3} $ 
units of $(2\pi/\tau)$. Restricted SOC is characterized by sharp decreasing of 
intensity on two orders for c and d cases! One can see color noises nearby 0.1 
and 0.2 f.u. for $\mu_1$ in state Ib.

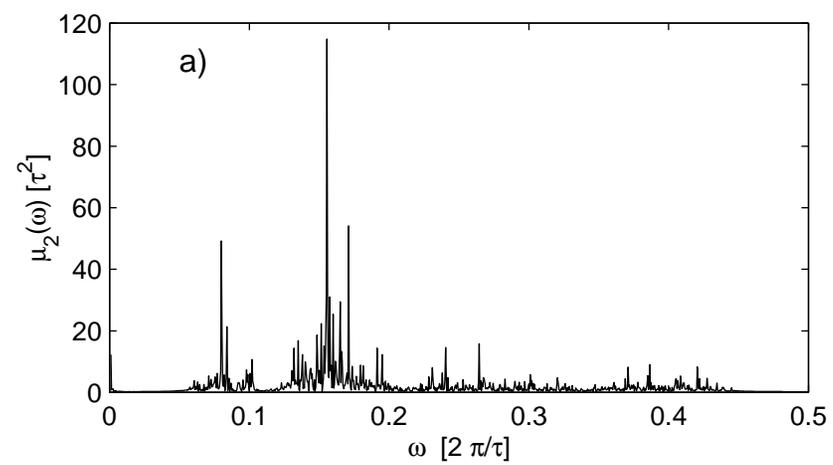
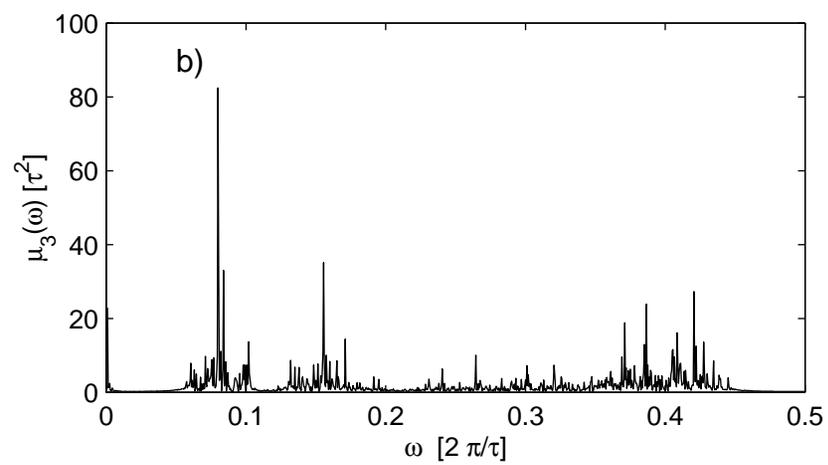
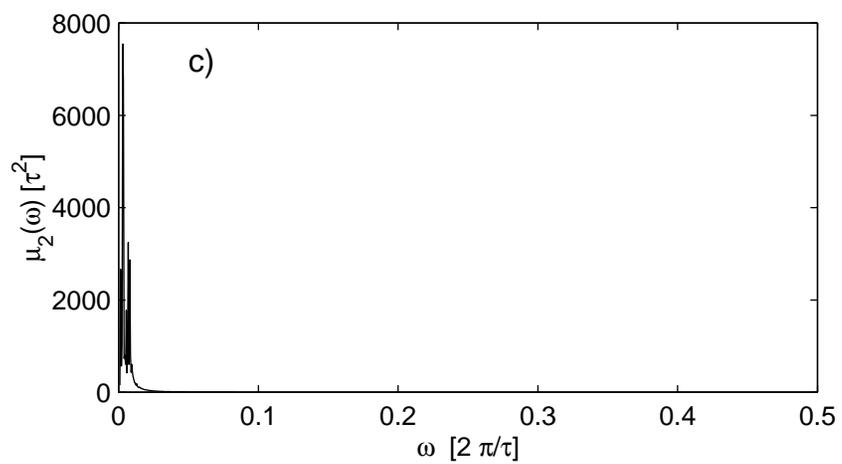
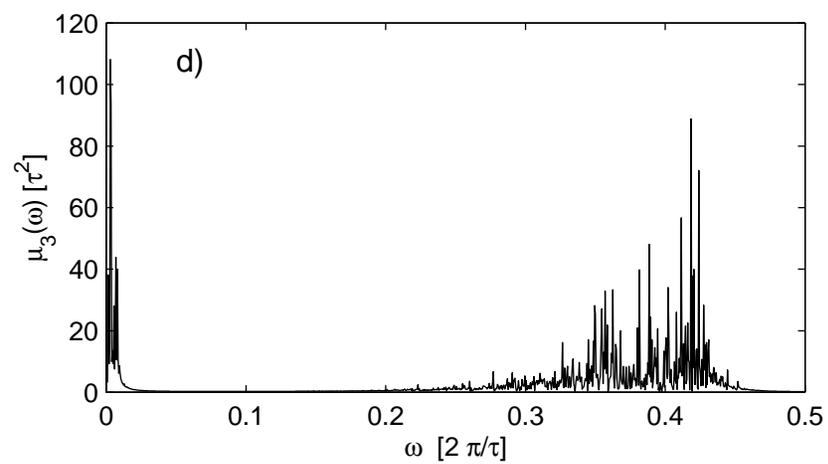
Figure 5. The spectra of two memory functions $\mu_2$ and
$\mu_3$: (a,b) -before the strong EQ,  (c,d)- and during  the strong EQ.  One 
can  observe color noises in 
cases a,b and d. Fractal-like spectrum  on ultra-low 
frequencies is appreciable in addition 
to cases c and d. The spectra for states Ib and I are sharply
different from  each other both to intensity and to  spectral peaks positioning.

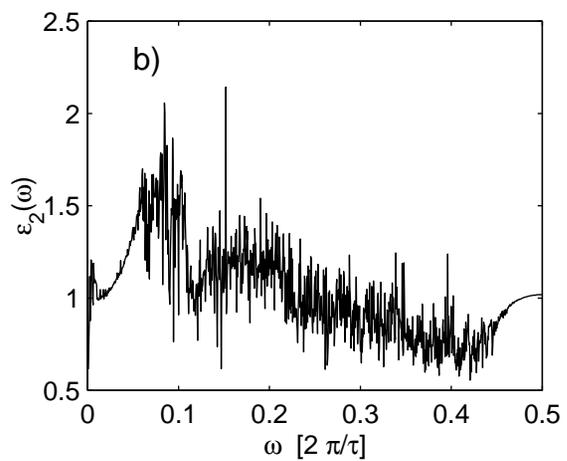
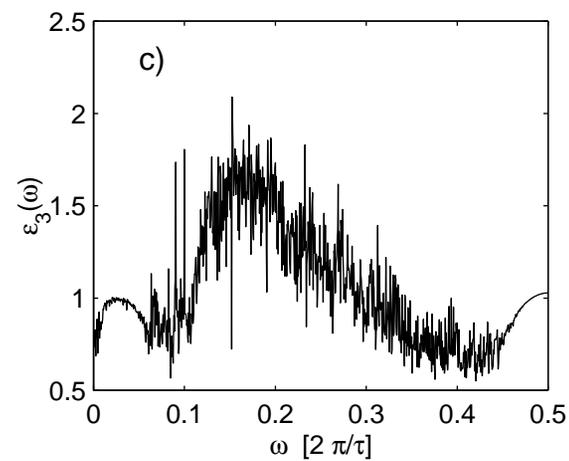
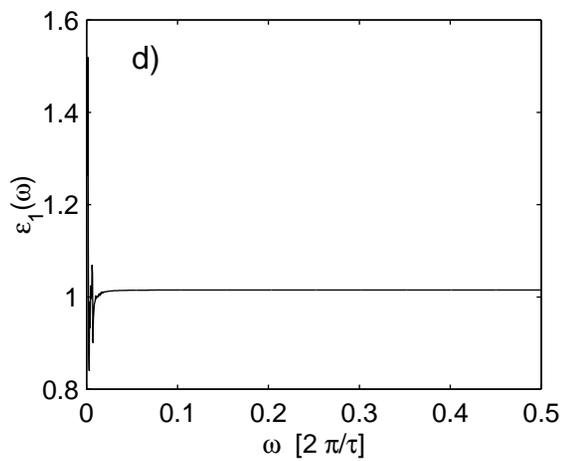
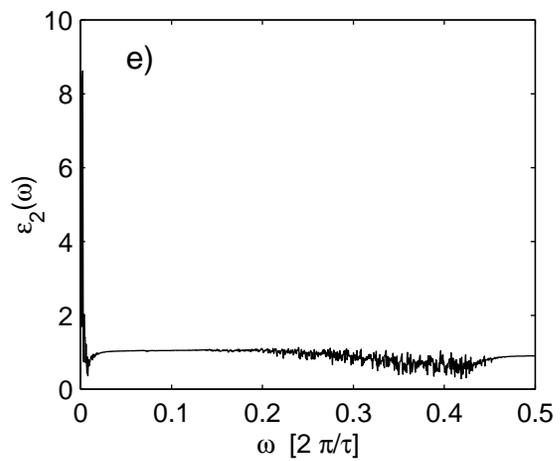
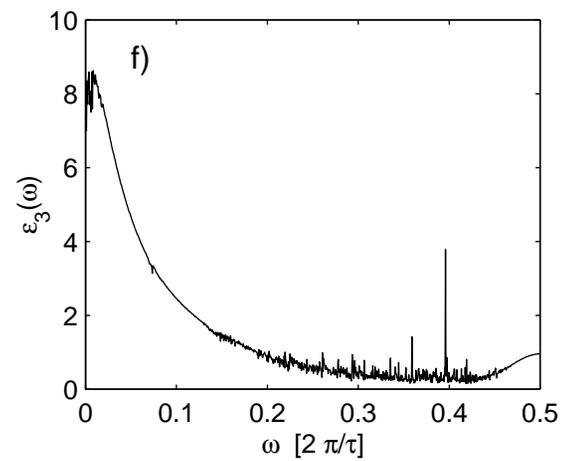
Figure 6. Frequency spectra of the first three points of
non-Markovity parameters $\epsilon_1$ 
$\epsilon_2$, $\epsilon_3$: (a,b,c)- before the strong EQ,
(d,e,f)- during the strong EQ. Markov and quasi-Markov behavior of seismic 
signals is observed only for $\epsilon_1$ in state Ib. All remaining cases 
(b,c,d and d) relate to non-Markov processes. Strong non-Markovity is typical 
for cases b , c (state Ib) and for case d (state I). In behavior of 
$\epsilon_2(\omega)$ and 
$\epsilon_3(\omega)$ one can see a transition from quasi-Markovity (at low 
frequencies) to strong non-Markovity (at high frequencies).

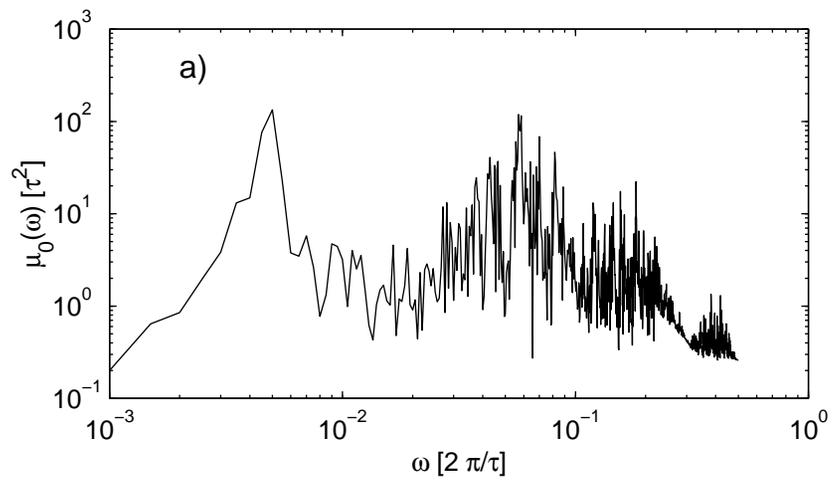
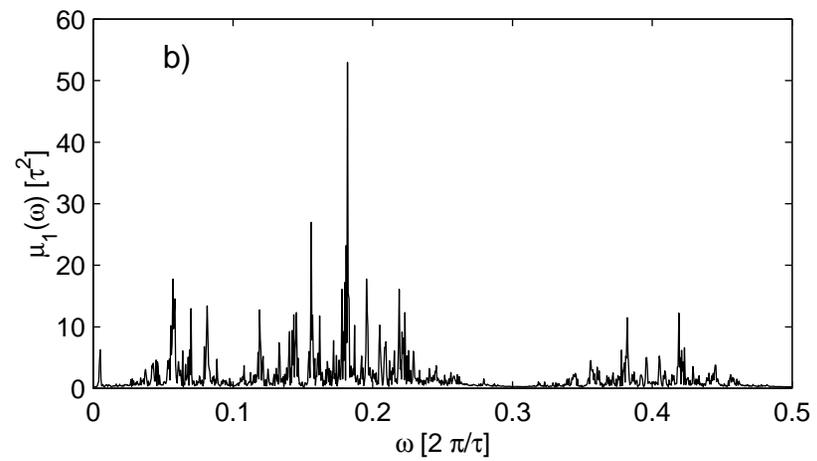
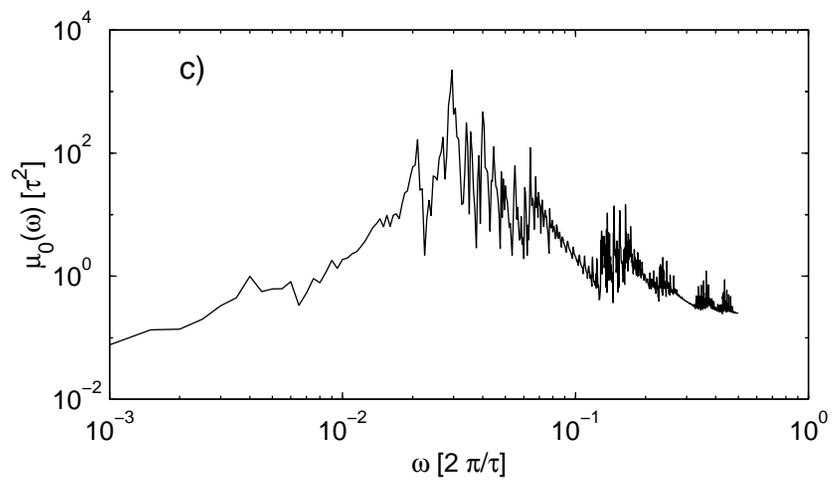
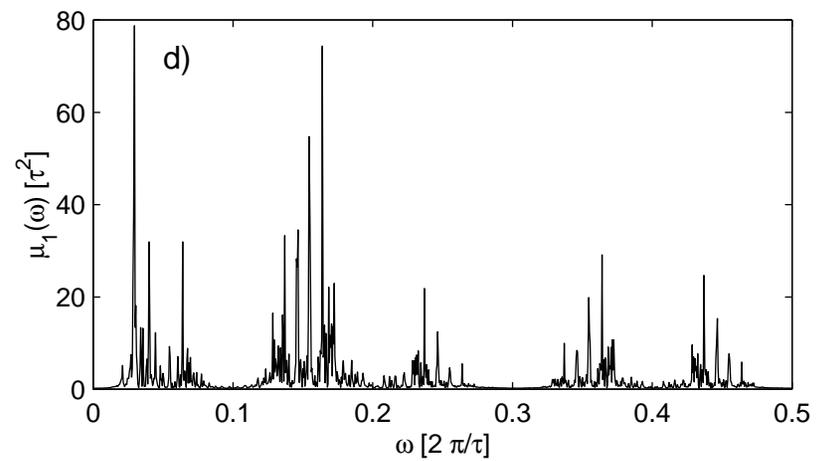
Figure 7. The power spectra for the two first memory functions $\mu_0$ and
$\mu_1$: (a,b) -during weak EQ,  (c,d)- during technogenic explosion. In 
cases Ib and I the spectra 
are characterized by strong differences 
especially on ultra low frequencies. They have very low intensity for $\mu_0$ on 
low frequencies (cases a and c) and color like behavior for $\mu_1$ for states 
II and III(cases b and d). Unexpected peaks exist in system III in LFR. 
Color and intensity distribution of the spectra is different for states II and 
III.

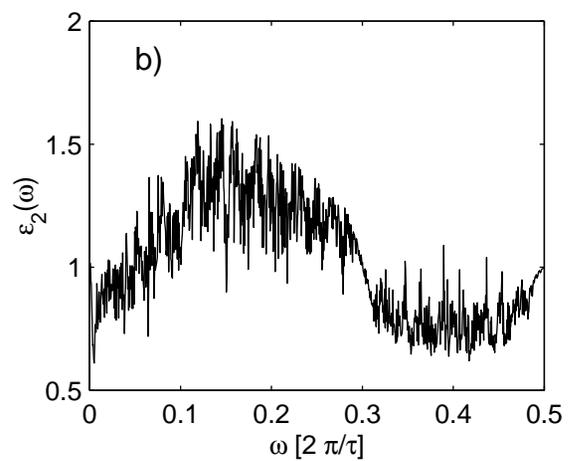
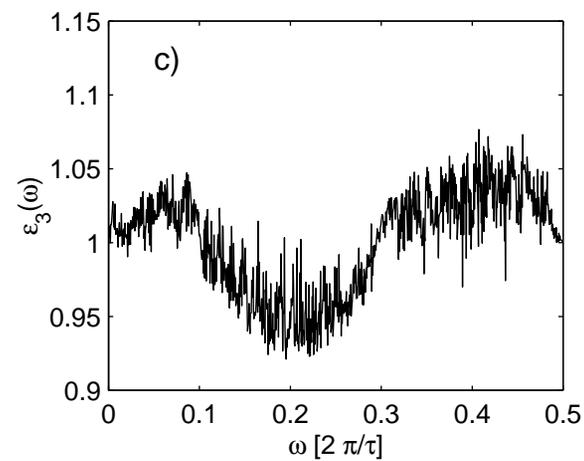
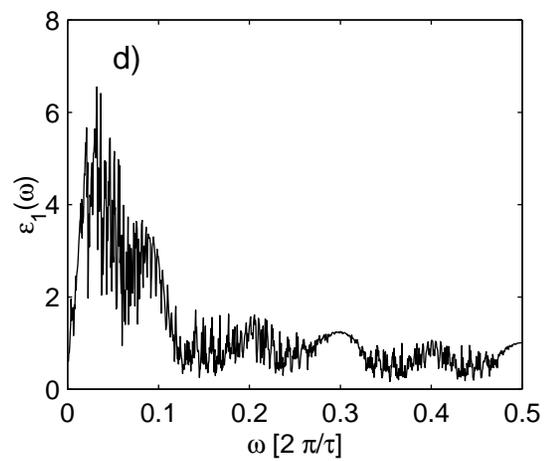
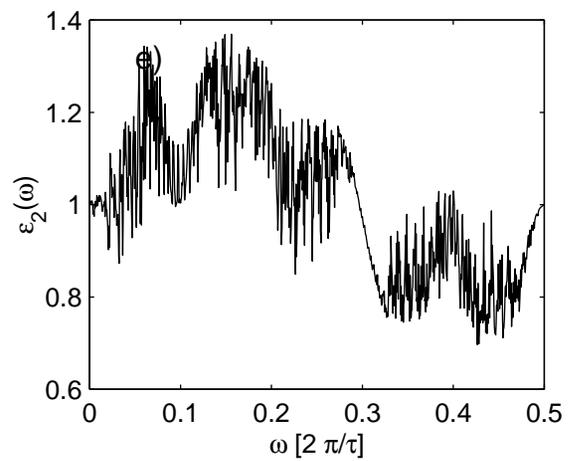
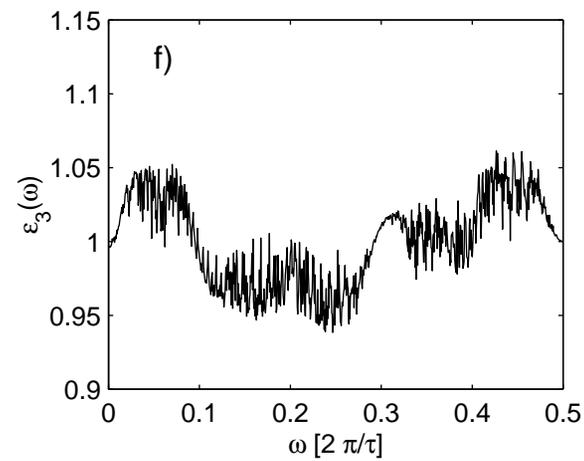
Figure 8. The frequency spectra of the first three points in statistical
spectrum of non-Markovity parameters
 $\epsilon_1$, $\epsilon_2$, $\epsilon_3$: (a,b,c)- during weak EQ,
(d,e,f)- during TE . All spectra are characterized by strong expressed non-
Markovity ( $\epsilon_i \sim 1$)  for the whole frequency range. Weak quasi-
Markovity is 
observed near zero frequency  for cases a and d 
($\epsilon_1$ vary from 0.5 up to 6.5). A noticeable difference for states II 
and 
III exists in behavior 
$\epsilon_1(\omega)$ in point $\omega= 0$. Due to this fact, one can develop  
reliable approach to differentiation between weak EQ's and underground TE's.

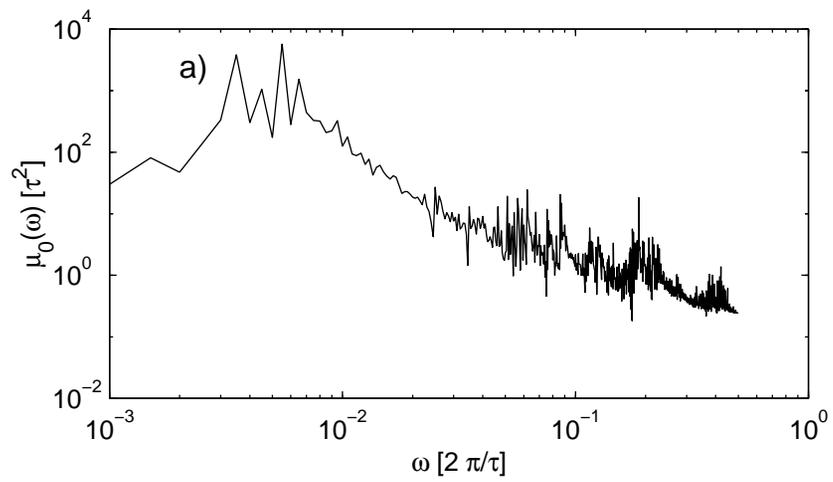
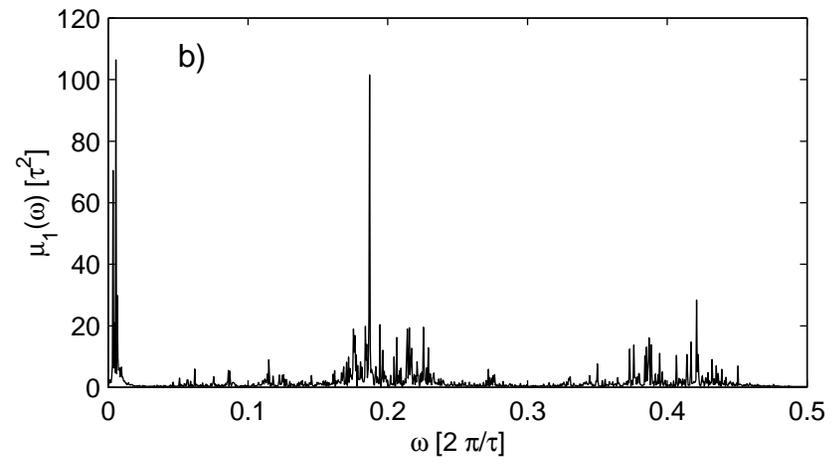
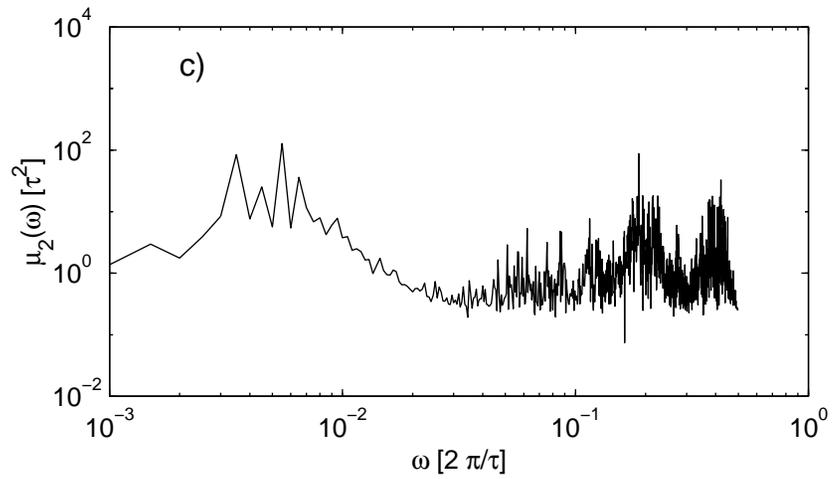
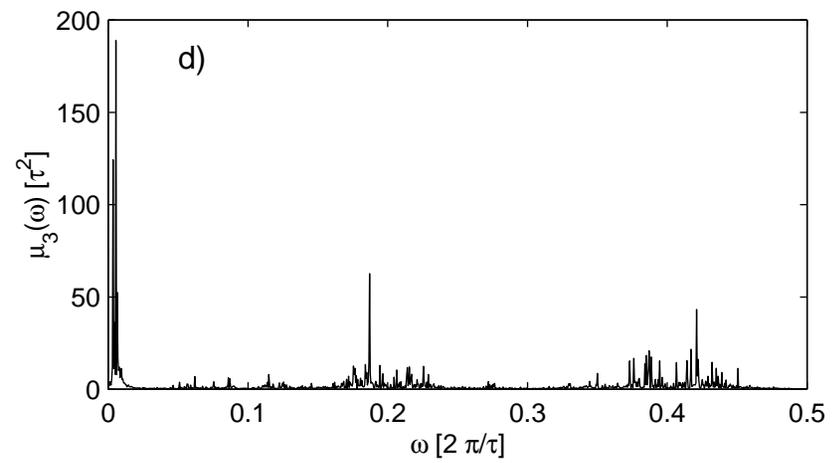
Figure 9. The power spectra of memory functions  
$\mu_0(\omega)$, $\mu_1(\omega)$, $\mu_2(\omega)$ and $\mu_3(\omega)$ 
for the calm state of the Earth before explosion. All functions $\mu_i(\omega)$, 
$ i=0,1,2,3 $ have approximately similar fractal behavior with restricted SOC 
and color noises close to 0.2 and 0.4 f.u. The maximum of intensity emerges 
close to the frequency  $4*10^-3 f.u. $ .  A slight change and redistribution of 
intensity of power spectra occur with the increase of order of memory function.

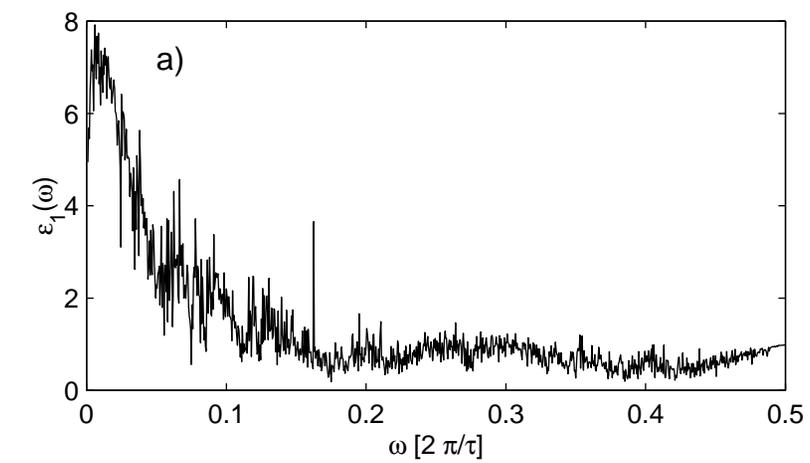
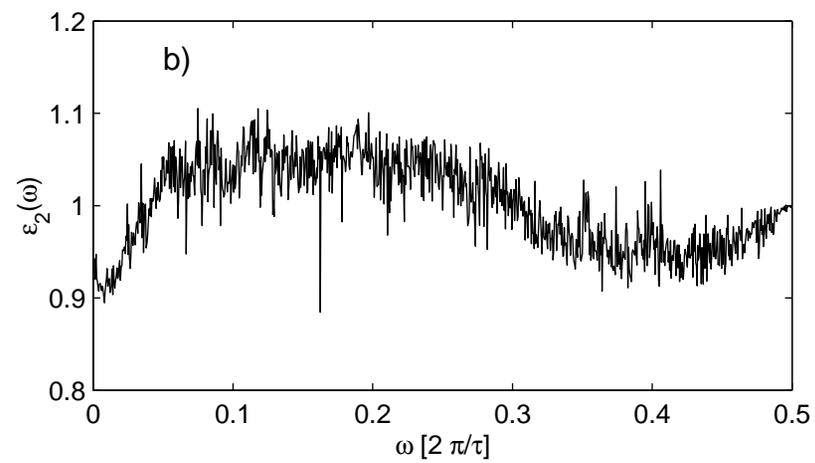
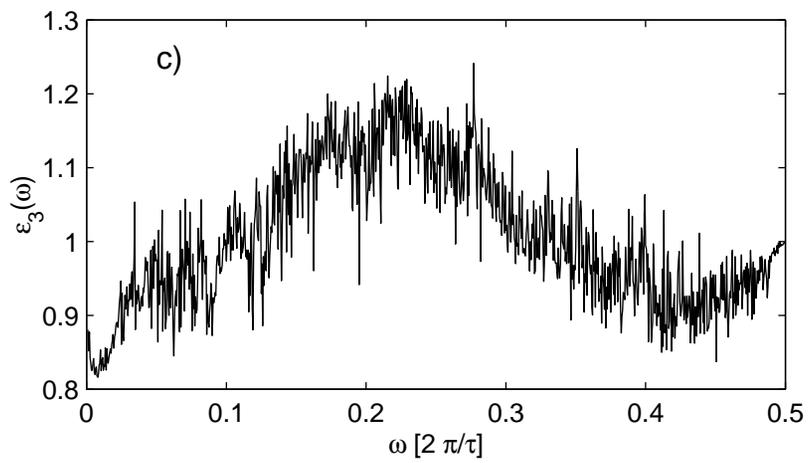
Figure 10. The power spectra of the first three points in statistical
spectrum of non-Markovity parameter
$\epsilon_1$, $\epsilon_2$, $\epsilon_3$ for calm state of 
the Earth before explosion(IV). Due to similar frequency behavior of all memory 
functions $\mu_i(\omega)$  the functions  $\epsilon_i(\omega)$ , i=1,2 and  3  
have
approximately similar frequency behavior and therefore demonstrate
strong non-Markovity on all levels. The initial parameter $\epsilon_1(\omega)$ 
is  non-Markovian with the exception of slight quasi-Markovity close to low 
frequencies below 0.1 f.u. 
As a result of this the possibility appears for forecasting the strong EQ's by  
registration of disappearance of strong non-Markovity and appearance of 
pronounced Markov time effects.

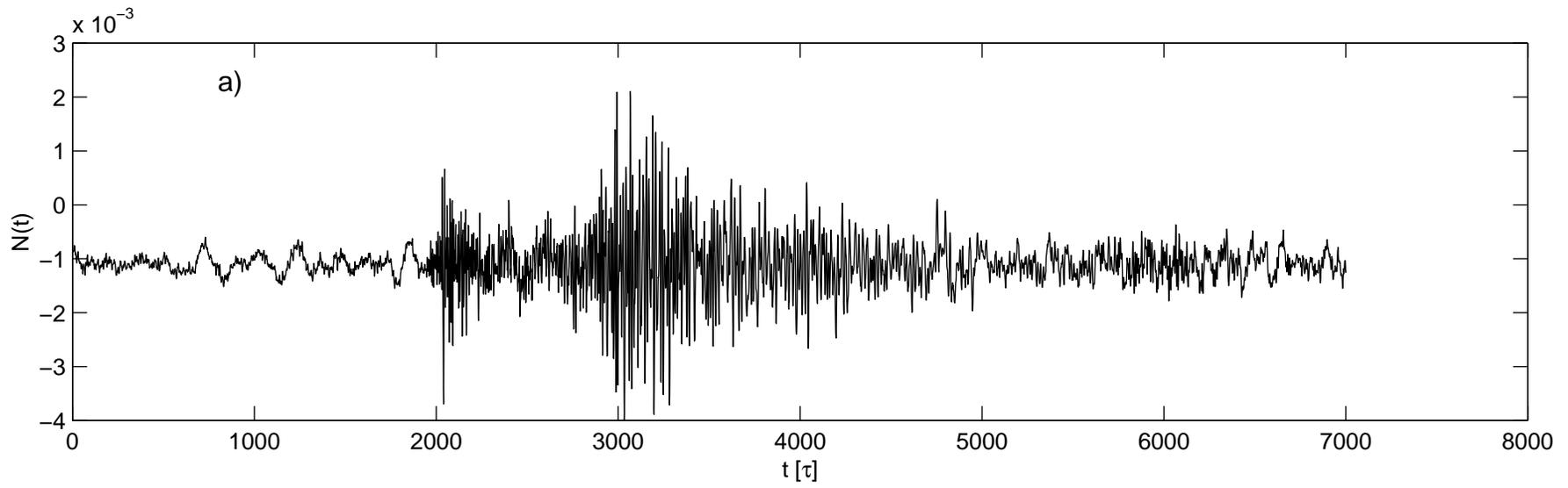
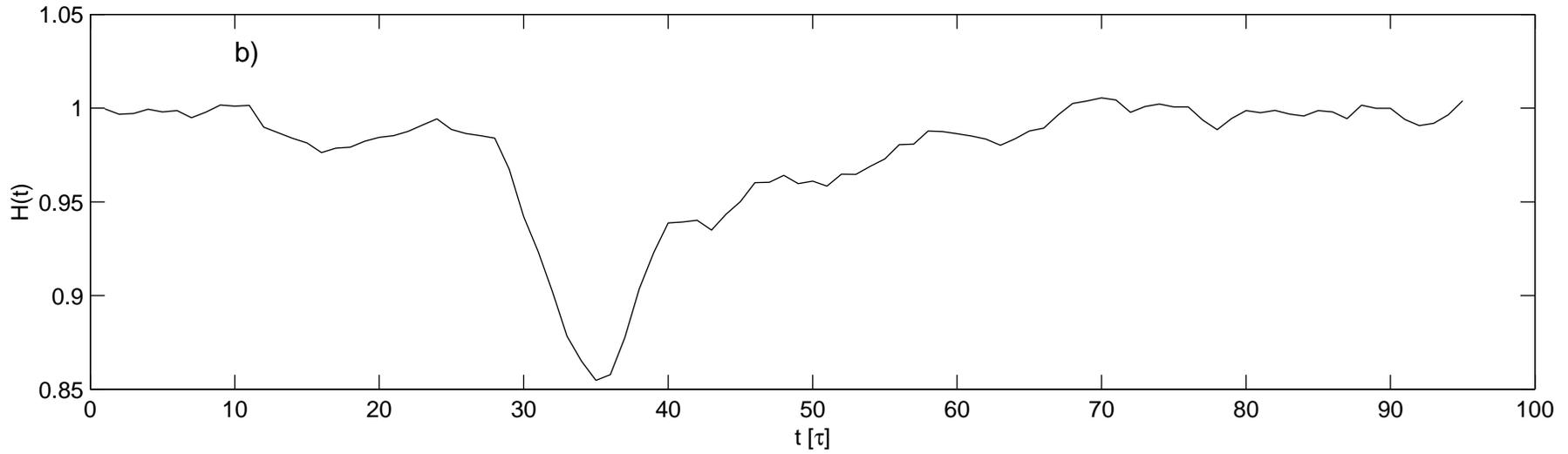
Figure 11. The typical temporal behavior of the Hurst exponent $H(t)$
calculated for  EQ's. One can see sharp decreasing of $H(t)$ on
15 \%  during  EQ. After that a gradual restore of the Hurst exponent $H(t)$ to 
normal value  $\approx 1$ 
takes place.

Figure 12. The comparative analysis of the Hurst exponent $H(t)$ behavior during 
the weak EQ (a,c) and  for the TE (b,d). During the weak EQ's one can see sharp 
decreasing of 
$H(t)$ on 15 \%  and almost  90 \%  
during the
TE. These observations enable us to develop a new approach to differentiation 
the TE's 
from weak EQ's.

\end{document}